\documentclass{ws-ijmpa}
\usepackage[super,compress]{cite}
\usepackage{graphicx}
\begin{document}
\markboth{K. A. Boshkayev et al.}{Geodesics in the field of rotating deformed gravitational source}

\catchline{}{}{}{}{}
%
%%%%%%%%%%%%%%%%%%%%%%%%%%%%%%%%%%%%%%%%%%%%%%%%%%%%%%%%%%%%%%%%%%%%

\title{Geodesics in the field of a rotating deformed gravitational source}

\author{K. A. Boshkayev,$^{1,2,^*}$  H. Quevedo,$^{2,3}$  M. S. Abutalip,$^{1}$\\ Zh. A. Kalymova$^{1}$ and Sh. S. Suleymanova$^{1}$}

\address{$^1$Department of Physics and Technology, Al-Farabi Kazakh National University,\\
Al-Farabi avenue 71, Almaty, 050040, Kazakhstan\\
$^2$International Center for Relativistic Astrophysics Network,\\
Piazza della Repubblica 10, Pescara, I-65122, Italy\\
$^3$Instituto de Ciencias Nucleares, Universidad Nacional Aut\'onoma de M\'exico, AP 70543, M\'exico, DF 04510, Mexico\\
$^*$kuantay@mail.ru}

\maketitle

\begin{history}
\received{Day Month Year}
\revised{Day Month Year}
\end{history}

\begin{abstract}
We investigate equatorial geodesics in the gravitational field of a rotating and deformed source described by the approximate Hartle-Thorne metric. In the case of 
massive particles, we derive within the same approximation analytic expressions for the orbital angular velocity, the specific angular momentum and energy, and the radii of marginally stable and  marginally bound circular orbits. Moreover, we calculate the orbital angular velocity and the radius of lightlike circular geodesics. We study numerically the frame dragging effect and the influence of the quadrupolar deformation of the source on the motion of test particles. We show that the effects originating from the rotation can be balanced by the effects due to the oblateness of the source. 
 
\end{abstract}

\ccode{PACS numbers: 04.20.-q; 04.20.Cv; 04.25.-g; 04.25.Nx; 04.25.D}

%\tableofcontents

\section{Introduction}	

In general, observed astrophysical objects  are characterized by a non-spherically symmetric distribution of mass and by rotation. In many cases, like ordinary planets and satellites, it is possible to neglect the deviations from spherical symmetry and the frame dragging effect, so that the gravitational field can be described
by the exterior Schwarzschild solution. In fact, the three classical tests of general relativity make use of the Schwarzschild spacetime in order to describe
gravity within the Solar system\cite{mtw}. In the case of strong gravitational fields, however, the deviation from spherical symmetry and the rotation become important and must be taken into account, at least to some extent. 

The first metric describing the exterior field of a slowly rotating slightly deformed object was found by Hartle and Thorne\cite{H,HT} in 1968. 
Alternative methods were proposed independently by Fock and Abdildin\cite{fock59,abd85} and Sedrakyan and Chubaryan\cite{sc68}. Only recently, it was
shown that in fact all these approaches are equivalent from a mathematical point of view\cite{bqr12,bqtz15}. At 
the level of the interpretation of the parameters entering the metric used in each approach, certain differences can appear which could make a particular 
approach more suitable for the investigation of certain problems. 

For the purpose of the present work, it is convenient to use the Hartle-Thorne
formalism which leads to an approximate metric describing, up to the first order in the quadrupole and the second order in the angular momentum, 
the exterior gravitational field of a rotating deformed object.  We will use in this work the Hartle-Thorne metric in the form presented  
by Bini et al.\cite{Bini2009} which in geometrical units is given by
\begin{eqnarray}\label{ht1}
ds^2&=&-\left(1-\frac{2{ M }}{r}\right)\left[1+2k_1P_2(\cos\theta)+2\left(1-\frac{2{ M}}{r}\right)^{-1}\frac{J^{2}}{r^{4}}(2\cos^2\theta-1)\right]dt^2 \nonumber\\
&&+\left(1-\frac{2{M}}{r}\right)^{-1}\left[1-2k_2P_2(\cos\theta)-2\left(1-\frac{2{M}}{r}\right)^{-1}\frac{J^{2}}{r^4}\right]dr^2 \nonumber\\
&&+r^2[1-2k_3P_2(\cos\theta)](d\theta^2+\sin^2\theta d\phi^2)-4\frac{J}{r}\sin^2\theta dt d\phi\ ,
\end{eqnarray}
\noindent where
\begin{eqnarray}\label{ht2}
k_1&=&\frac{J^{2}}{{M}r^3}\left(1+\frac{{M}}{r}\right)-\frac{5}{8}\frac{Q-J^{2}/{M}}{{M}^3}Q_2^2\left(\frac{r}{{M}}-1\right)\ , \quad k_2=k_1-\frac{6J^{2}}{r^4}\ , \nonumber\\
k_3&=&k_1+\frac{J^{2}}{r^4}-\frac{5}{4}\frac{Q-J^{2}/{M}}{{M}^2r}\left(1-\frac{2{M}}{r}\right)^{-1/2}Q_2^1\left(\frac{r}{M}-1\right)\ .
\end{eqnarray}
Here $P_{2}(\cos\theta)=\frac{1}{2}(3\cos^{2}\theta-1)$ is Legendre polynomial of the first kind, $Q_l^m$ are the associated Legendre functions of the second kind determined as
\begin{eqnarray}\label{}
Q_{2}^{1}(x)&=&(x^{2}-1)^{1/2}\left[\frac{3x}{2}\ln\frac{x+1}{x-1}-\frac{3x^{2}-2}{x^{2}-1}\right], \nonumber\\
Q_{2}^{2}(x)&=&(x^{2}-1)\left[\frac{3}{2}\ln\frac{x+1}{x-1}-\frac{3x^{3}-5x}{(x^{2}-1)^2}\right],
\end{eqnarray}
and the constants ${M}$, ${J}$ and ${Q}$ are the total mass, angular momentum and mass quadrupole moment of the rotating object, respectively 
(for more details see Refs.~\refcite{H} and \refcite{HT}).\footnote{We note here that the quadrupole parameter $Q$ is related to the mass quadrupole moment defined by Hartle and Thorne\cite{HT} through $Q=2J^2/M-Q_{HT}$.}

The approximate Kerr metric\cite{K} in Boyer-Lindquist coordinates can be obtained from the above Hartle-Thorne metric after setting
\begin{equation}\label{tr1}
J=-Ma,\quad Q={J}^2/{M},
\end{equation}
and making a coordinate transformation implicitly given by
\begin{eqnarray}\label{tr2}
t&=&t ,\quad \phi=\phi ,\nonumber\\
r&=&R+\frac{a^2}{2R}\left[\left(1+\frac{2M}{R}\right)\left(1-\frac{M}{R}\right)-\cos^2\Theta\left(1-\frac{2M}{R}\right)\left(1+\frac{3M}{R}\right)\right]\ , \nonumber\\
\theta&=&\Theta+\frac{a^2}{2R^2}\left(1+\frac{2M}{R}\right)\sin\Theta\cos\Theta.
\end{eqnarray}

The Kerr metric is important to investigate the physical processes taking place around rotating black holes, i.e., the source with probably the strongest possible gravitational field. The role of rotation is essential in the physics of accretion disks and energy extraction from a black hole. Moreover, depending on the direction of the rotation, the radius of the accretion disk can be larger or smaller with respect to the Schwarzschild case. The situation changes when one involves compact objects such as white dwarfs, neutron stars and quark stars as they have additional parameters to be taken into account. The combination of the strong field with the quadrupolar deformation of the source plays a pivotal role when one considers the motion of test particles. 

There exist many exact solutions that include a quadrupole parameter. The importance of the quadrupole moment in the astrophysical context has been emphasized  in several works\cite{1989GReGr..21.1047Q, quevedo2011, Bini2009, 2012CQGra..29n5003B, 2012ApJ...756...82P, bini2013}. However, most analysis  must be performed
numerically due to the complexity of the exact metrics. The advantage of considering the Hartle-Thorne approximate solution is that several physical quantities can
be calculated analytically which facilitates their study. We will prove below that this is possible for a particular set of geodesics. 
In this work, we are interested in studying the motion of test particles in the Hartle-Thorne spacetime. Therefore, we will perform both analytical and numerical 
analysis of the timelike and lightlike geodesic equations. In particular, we are interested in comparing the effects of the quadrupole and angular momentum parameters
within the approximation allowed by the Hartle-Thorne metric.

\section{Geodesic Equations: Analytic Results}
%\subsection{Normalization Condition}

In this work, we will make use the timelike normalization condition
$ U^{\alpha}U_{\alpha}=-1$ 
which for equatorial circular geodesics  is equivalent to
\begin{equation}
g_{tt}(U^t)^2+2g_{t\phi}U^{t}U^{\phi}+g_{\phi\phi}(U^{\phi})^2=-1 \ , \quad 
U^t=\frac{dt(s)}{ds},\quad U^{\phi}=\frac{d\phi(s)}{ds}.
\end{equation}
Sometimes the following convenient notations are used for the four-velocity of equatorial circular geodesics
\begin{equation}
U^t=\Gamma,\quad U^{\phi}=\Gamma\zeta,
\end{equation}
\noindent where $\Gamma$ is the normalization factor and $\zeta$ is the orbital angular velocity.
%%%%%%%%%%%%%%%%%%%%%%%%%%%%%%%%%%%%%%%%%%%%%%%%%%%%%%%%%%%%%%%%%%%%%%%%%%%%%%%%%%%%%%%%%%%%%%%%%%%%%%%%%%%%%%%%%%%%%%%%%%%%%%%%%%%%%%%%%%%%%%%%%%%%%%%%%%%%%%%%%%%%%%%%%%%%
%\subsection{Equations For the Equatorial Circular Geodesics}

Using the fact that the Hartle-Thorne solution possesses two Killing vector fields $\partial_t$ and $\partial_\phi$, which determine two constants of motion, 
from the geodesic equations for equatorial circular orbits ($\theta=\pi/2$ and $ r=const)$, we obtain 
\begin{equation}
g_{tt,r}(U^t)^2+2g_{t\phi,r}U^{t}U^{\phi}+g_{\phi\phi,r}(U^{\phi})^2=0,
\end{equation}
\noindent where a comma indicates partial differentiation. Then, a straightforward computation yields 
\begin{eqnarray}
&&(r-2M)\left\{\frac{M}{r^3}\left[\frac{dt(s)}{ds}\right]^2-\left[\frac{d\phi(s)}{ds}\right]^2\right\}-\frac{2M^2(r-2M)}{r^3}\frac{dt(s)}{ds}\frac{d\phi(s)}{ds}j \nonumber\\
&&\qquad\qquad\qquad+\left\{A_{1}(r)\left[\frac{dt(s)}{ds}\right]^2+A_{2}(r)\left[\frac{d\phi(s)}{ds}\right]^2\right\}j^2 \nonumber\\
&& \qquad\qquad\qquad\qquad\qquad\qquad +\left\{A_{3}(r)\left[\frac{dt(s)}{ds}\right]^2+A_{4}(r)\left[\frac{d\phi(s)}{ds}\right]^2\right\}q = 0 ,
\end{eqnarray}
\noindent where we introduced the dimensionless quantities $j=J/M^2$ and $q=Q/M^3$ and new functions defined as
\begin{eqnarray}
A_{1}(r)&=&-\frac{16M^5-42M^4r-30M^3r^2+25M^2r^3+30Mr^4-15r^5}{8Mr^5}-B_{1}(r), \nonumber \\
A_{2}(r)&=&\frac{48M^6-64M^5r+12M^4r^2-70M^2r^4+15Mr^5+15r^6}{8Mr^4}-B_{2}(r),\nonumber \\
A_{3}(r)&=&-\frac{5(6M^4+6M^3r-5M^2r^2-6Mr^3+3r^4)}{8Mr^4}+B_{1}(r),\nonumber \\
A_{4}(r)&=&\frac{5(14M^2-3Mr-3r^2)}{8M}+B_{2}(r),\nonumber \\
B_{1}(r)&=&\frac{15(4M^3-3Mr^2+r^3)}{16M^2r^2}\ln\frac{r}{r-2M},\nonumber \\
B_{2}(r)&=&\frac{15(4M^3-6M^2r+r^3)}{16M^2}\ln\frac{r}{r-2M}.
\end{eqnarray}
This expansion in terms of the quadrupole and angular momentum parameters can be used to derive analytical expressions for the parameters that 
characterize the orbits of the test particles. 

%%%%%%%%%%%%%%%%%%%%%%%%%%%%%%%%%%%%%%%%%%%%%%%%%%%%%%%%%%%%%%%%%%%%%%%%%%%%%%%%%%%%%%%%%%%%%%%%%%%%%%%%%%%%%%%%%%%%%%%%%%%%%%%%%%%%%%%%%%%%%%%%%%%%%%%%%%%%%%%%%%%%%%%%%%%%
\subsection{The orbital angular velocity for test particles $(\zeta=U^{\phi}/U^{t})$}
Starting from the $r$ component of the geodesic equation and using the fact that $r$ and $\theta$ are both constant on circular equatorial orbits, one easily derives the following expression for the angular velocity:
\begin{equation}
\zeta(r)=\frac{-g_{t\phi,r}\pm\sqrt{(g_{t\phi,r})^2-g_{tt,r}g_{\phi\phi,r}}}{g_{\phi\phi,r}}\ .
\end{equation}
Performing an analysis similar to the one carried out in the previous section, we finally obtain
\begin{equation}
\zeta(r)=\pm\zeta_{0}(r)\left[1\mp F_{1}(r)j+F_{2}(r)j^2+F_{3}(r)q\right],
\end{equation}
\noindent where
\begin{eqnarray}
\zeta_{0}(r)&=&\frac{M^{1/2}}{r^{3/2}}, \quad F_{1}(r)=\frac{M^{3/2}}{r^{3/2}}, \nonumber\\
F_{2}(r)&=&[16M^2r^4(r-2M)]^{-1}[48M^7-80M^6r+4M^5r^2+42M^4r^3\nonumber\\
&&-40M^3r^4-10M^2r^5-15Mr^6+15r^7]-F(r), \nonumber\\
F_{3}(r)&=&-\frac{5(6M^4-8M^3r-2M^2r^2-3Mr^3+3r^4)}{16M^2r(r-2M)}+F(r), \nonumber\\
F(r)&=&\frac{15(r^3-2M^3)}{32M^3}\ln\frac{r}{r-2M}.
\end{eqnarray}

This expression for the orbital angular velocity of a test particle at the equatorial plane can be used to determine the mass shedding limit of the source in general relativity\cite{boshkayev2013, 2010ApJ...714..748T}. Moreover, in X-Ray astronomy, the orbital angular velocity is associated with the upper frequency of the quasiperiodic oscillations\cite{A2003, 2010ApJ...714..748T, abr2004, boshkayevqpos2014}. The analytic expression obtained here leads to results that are in agreement 
with those obtained by using pure numerical methods. 
%&&&&&&&&&&&&&&&&&&&&&&&&&&&&&&&&&&&&&&&&&&&&&&&&&&&&&&&&&&&&&&&&&&&&&&&&&&&&&&&&&&&&&&&&&&&&&&&&&&&&&&&&&&&&&&&&&&&&&&&&&&&&&&&&&&&&&&&&&&&&&&&&&&&&&&&&&&
\subsection{The orbital angular velocity for photons $(\zeta_{ph}=U^{\phi}/U^{t})$}
In the case of lightlike geodesics, we can use the expression for the norm of the corresponding 4-velocity 
to calculate its components. Evaluating $U^{\phi}$ and $U^{t}$ directly from the Hartle-Thorne line element, we then obtain for the orbital angular velocity:
\begin{equation}
\zeta_{ph}(r)=\pm\zeta_{ph0}(r)\left[1\mp W_{1}(r)j+W_{2}(r)j^2+W_{3}(r)q\right],
\end{equation}
\noindent where
\begin{eqnarray}
\zeta_{ph0}(r)&=&\frac{(r-2M)^{1/2}}{r^{3/2}} , \quad W_{1}(r)=\frac{2M^{2}}{r^{3/2}(r-2M)^{1/2}} , \nonumber\\
W_{2}(r)&=&\frac{24M^6+12M^5r-8M^4r^2+15M^3r^3-10M^2r^4-30Mr^5+15r^6}{8Mr^4(r-2M)}+W(r) , \nonumber\\
W_{3}(r)&=&-\frac{5(3M^3-2M^2r-6Mr^2+3r^3)}{8Mr(r-2M)}-W(r) , \nonumber\\
W(r)&=&\frac{15(M^2+Mr-r^2)}{16M^2}\ln\frac{r}{r-2M} .
\end{eqnarray}
%%%%%%%%%%%%%%%%%%%%%%%%%%%%%%%%%%%%%%%%%%%%%%%%%%%%%%%%%%%%%%%%%%%%%%%%%%%%%%%%%%%%%%%%%%%%%%%%%%%%%%%%%%%%%%%%%%%%%%%%%%%%%%%%%%%%%%%%%%%%%%%%%%%%%%%%%%%%%%%%%%%%%%%%%%%%%%%%%%%%%%%%%%%%%%%%%%%%%%%%%%%%%%%%%%%%%%%%%%%%%%%%%%%%%%%%%
\subsection{The specific energy  per unit mass ($\varepsilon=-U_{t}$) for test particles}
The specific energy per unit mass $\varepsilon$ is usually used to estimate the radius of marginally (innermost) bound orbits of test particles. Thus, this radius determines the stable region which is essential for the formation of accretion disks.
Evaluating $U_{t}$ from the line element, one obtains
 \begin{equation}
\varepsilon=\frac{g_{tt}+\zeta g_{t\phi}}{\sqrt{-g_{tt}-2\zeta g_{t\phi}-\zeta^2g_{\phi\phi}}},
\end{equation}
\begin{equation}
\varepsilon=\varepsilon_{0}\left[1\mp H_{1}(r)j+H_{2}(r)j^2+H_{3}(r)q\right],
\end{equation}
\noindent where
\begin{eqnarray}
\varepsilon_{0}&=&\frac{r-2M}{r^{1/2}(r-3M)^{1/2}}, \quad H_{1}(r)=\frac{M^{5/2}}{r^{1/2}(r-2M)(r-3M)},\nonumber\\
H_{2}(r)&=&-[16M r^4(r-2M)(r-3M)^2]^{-1}[144M^8-144M^7r-28M^6r^2+122M^5r^3\nonumber\\
&&+184M^4r^4-685M^3r^5+610M^2r^6-225Mr^7+30r^8]+H(r), \nonumber \\
H_{3}(r)&=&\frac{5(r-M)(6M^3+20M^2r-21Mr^2+6r^3)}{16Mr(r-2M)(r-3M)}-H(r), \nonumber\\
H(r)&=&\frac{15r(8M^2-7Mr+2r^2)}{32M^2(r-3M)}\ln\frac{r}{r-2M}.
\end{eqnarray}
%%%%%%%%%%%%%%%%%%%%%%%%%%%%%%%%%%%%%%%%%%%%%%%%%%%%%%%%%%%%%%%%%%%%%%%%%%%%%%%%%%%%%%%%%%%%%%%%%%%%%%%%%%%%%%%%%%%%%%%%%%%%%%%%%%%%%%%%%%%%%%%%%%%%%%%%%%%%%%%%%%%%%%%%%%%%
\subsection{The specific angular momentum per unit energy  ($l=-U_{\phi}/U_{t}$)}
As we will see below, the specific angular momentum for test particles per unit energy $l$ is crucial for the determination of the marginally (innermost) stable orbits of test particles forming accretion disks. Calculating $U_{\phi}$ and $U_{t}$, we obtain the following analytic expression for $l$
\begin{equation}
l=-\frac{g_{t\phi}+\zeta g_{\phi\phi}}{g_{tt}+\zeta g_{t\phi}},
\end{equation}
\begin{equation}
l=\pm l_{0}\left[1\mp G_{1}(r)j+G_{2}(r)j^2+G_{3}(r)q\right],
\end{equation}
\noindent where
\begin{eqnarray}
l_{0}&=&\frac{M^{1/2}r^{3/2}}{r-2M}, \quad G_{1}(r)=\frac{M^{3/2}(3r-4M)}{r^{3/2}(r-2M)}, \nonumber \\
G_{2}(r)&=&[16M^2 r^4(r-2M)^2]^{-1}[96M^8-112M^7r-8M^6r^2+72M^5r^3\nonumber \\
&&-18M^4r^4-220M^3r^5+260M^2r^6-105Mr^7+15r^8]-G(r), \nonumber \\
G_{3}(r)&=&\frac{5(6M^4-22M^2r^2+15Mr^3-3r^4)}{16M^2 r(r-2M)}+G(r), \nonumber \\
G(r)&=&\frac{15(2M^3+4M^2r-4Mr^2+r^3)}{32M^3}\ln\frac{r}{r-2M}.
\end{eqnarray}
%%%%%%%%%%%%%%%%%%%%%%%%%%%%%%%%%%%%%%%%%%%%%%%%%%%%%%%%%%%%%%%%%%%%%%%%%%%%%%%%%%%%%%%%%%%%%%%%%%%%%%%%%%%%%%%%%%%%%%%%%%%%%%%%%%%%%%%%%%%%%%%%%%%%%%%%%%%%%%%%%%%%%%%%%%%%
\subsection{Radius of the photon, innermost bound and innermost stable orbits}
The normalization condition $P^{\alpha}P_{\alpha}=0$ gives the photon orbit, $r_{ph}$, where $P^{\alpha}$ is the photon four-momentum and for circular orbits $\alpha = t, \phi$. Indeed the normalization condition $P_{t}P^{t}+P_{\phi}P^{\phi}=0$ gives the orbital angular velocity for the photon $\zeta_{ph}$, but $\Gamma_{ph}$ remains arbitrary. To determine the photon orbit, $r_{ph}$, first one should use the above expression for $\zeta_{ph}$ and then evaluate the four-acceleration $a^{\alpha}$. For a circular geodesic $a^{\alpha}=0$, and only from this condition one can determine $r_{ph}$. Note, alternatively it is also convenient to use the condition $U^{t}=0$ to find $r_{ph}$. Moreover, in order to determine the radius of the innermost (marginally) bound circular orbits $r_{mb}$ one should use the condition $\varepsilon=1$. In addition, the condition $dl/dr=0$ allows one to find the radius of the innermost stable circular orbits, $r_{ms}$. Here we used the methods of perturbation theory and the results of these calculations are:
\begin{eqnarray}\label{rr}
r_{ph}&=&3M\left[1\pm\frac{2\sqrt{3}}{9}j+\left(\frac{1751}{324}-\frac{75}{16}\ln3\right)j^2+\left(-\frac{65}{12}+\frac{75}{16}\ln3\right)q\right],\\
r_{mb}&=&4M\left[1\mp\frac{1}{2}j+\left(\frac{8033}{256}-45\ln2\right)j^2+\left(-\frac{1005}{32}+45\ln2\right)q\right],\\
r_{ms}&=&6M\left[1\pm\frac{2}{3}\sqrt{\frac{2}{3}}j+\left(-\frac{251903}{2592}+240\ln\frac{3}{2}\right)j^2+\left(\frac{9325}{96}-240\ln\frac{3}{2}\right)q\right]. \qquad
\end{eqnarray}
%%&&&&&&&&&&&&&&&&&&&&&&&&&&&&&&&&&&&&&&&&&&&&&&&&&&&&&&&&&&&&&&&&&&&&&&&&&&&&&&&&&&&&&&&&&&&&&&&&&&&&&&&&&&&&&&&&&&
%%&&&&&&&&&&&&&&&&&&&&&&&&&&&&&&&&&&&&&&&&&&&&&&&&&&&&&&&&&&&&&&&&&&&&&&&&&&&&&&&&&&&&&&&&&&&&&&&&&&&&&&&&&&&&&&&&&&
\subsection{The epicyclic frequencies}
Finally, we mention that using the Hartle-Thorne line element it is possible to derive the radial and vertical fundamental frequencies.\cite{boshkayevqpos2014}
The application of these frequencies to the observed quasiperiodic oscillations from the low-mass X-ray binaries has been considered on the basis of the relativistic precession model.\cite{boshkayevqpos2015} All the epicyclic frequencies have been derived in previous works.\cite{A2003, 2010ApJ...714..748T, abr2004, stuchlik2014, stuchlik2015} With the method proposed in this work we obtained equivalent results after applying the redefinition $Q\rightarrow2J^2/M-Q$ or $q\rightarrow2j^2-q$. 
%%&&&&&&&&&&&&&&&&&&&&&&&&&&&&&&&&&&&&&&&&&&&&&&&&&&&&&&&&&&&&&&&&&&&&&&&&&&&&&&&&&&&&&&&&&&&&&&&&&&&&&&&&&&&&&&&&&&
\section{Equatorial Geodesics: Numerical Results}
\begin{figure}[t]
\centering
\begin{tabular}{lr}
\includegraphics[totalheight=6cm, width=6cm]{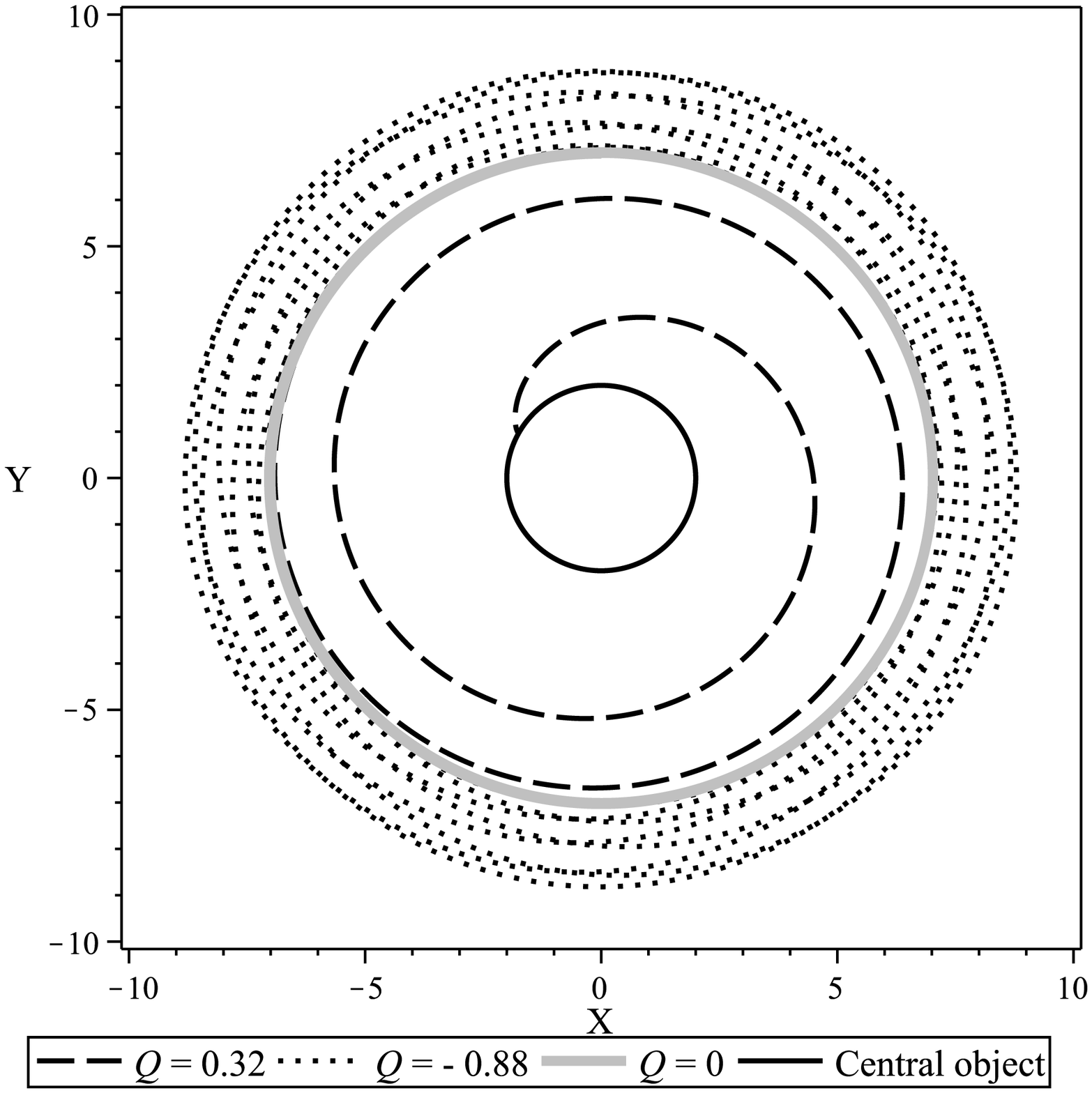} &
\includegraphics[totalheight=6cm, width=6cm]{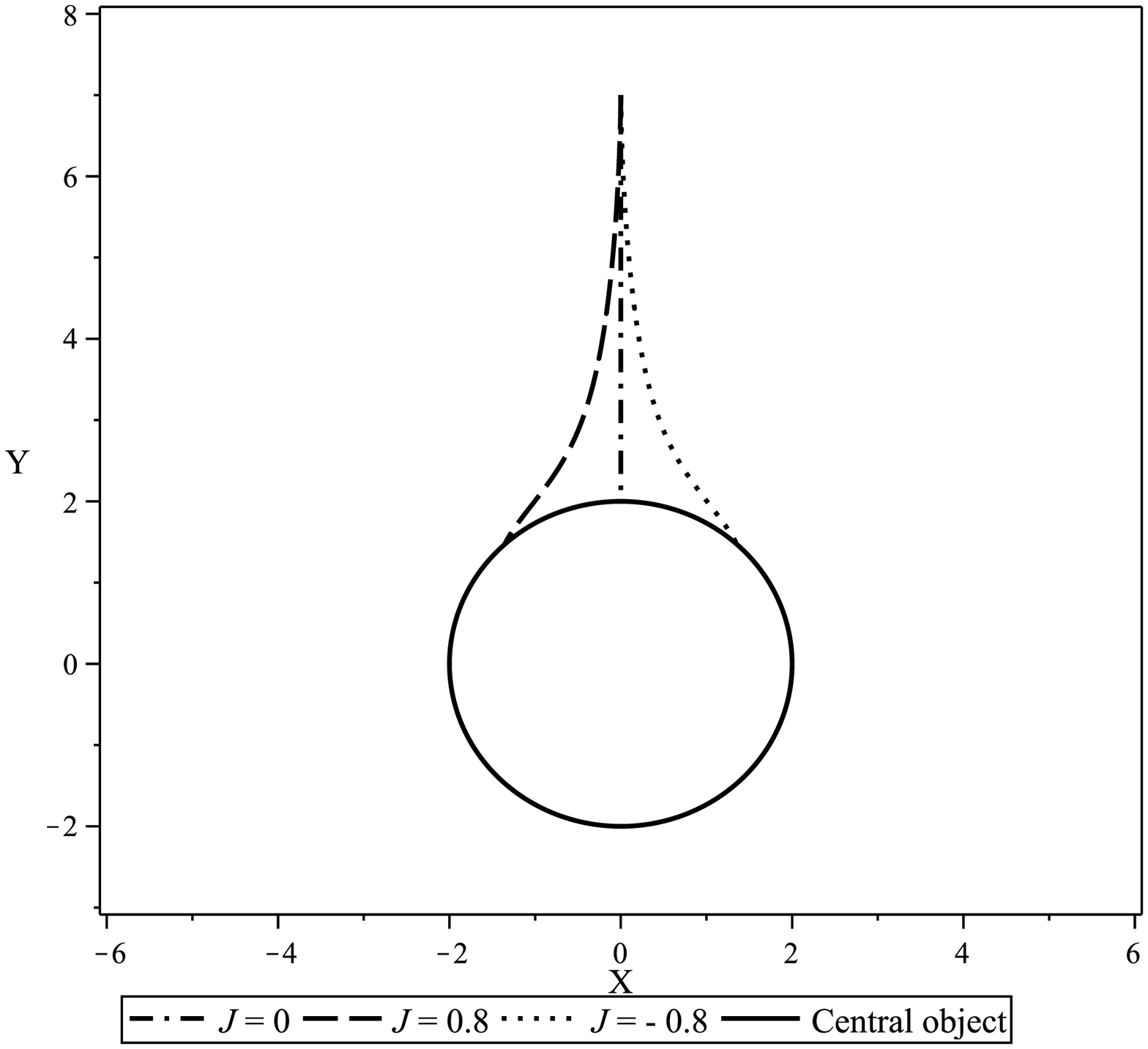}
\end{tabular}
\caption{(left) Motion of a test particle in the field of a static deformed object. 
The parameters of the system are $M=1,\quad J=0,\quad  r=7,\quad \phi=\pi/2, \quad \dot{r}=0, \quad \dot{\phi}=0.07145$. (right) Frame dragging. The parameters of the system are $M=1, \quad Q=0,\quad r=7, \quad \phi=\pi/2, \quad \dot{r}=0, \quad \dot{\phi}=0$.}\label{plotq}
\end{figure}

\begin{figure}[t]
\centering
\begin{tabular}{lr}
\includegraphics[totalheight=6cm,width=6cm]{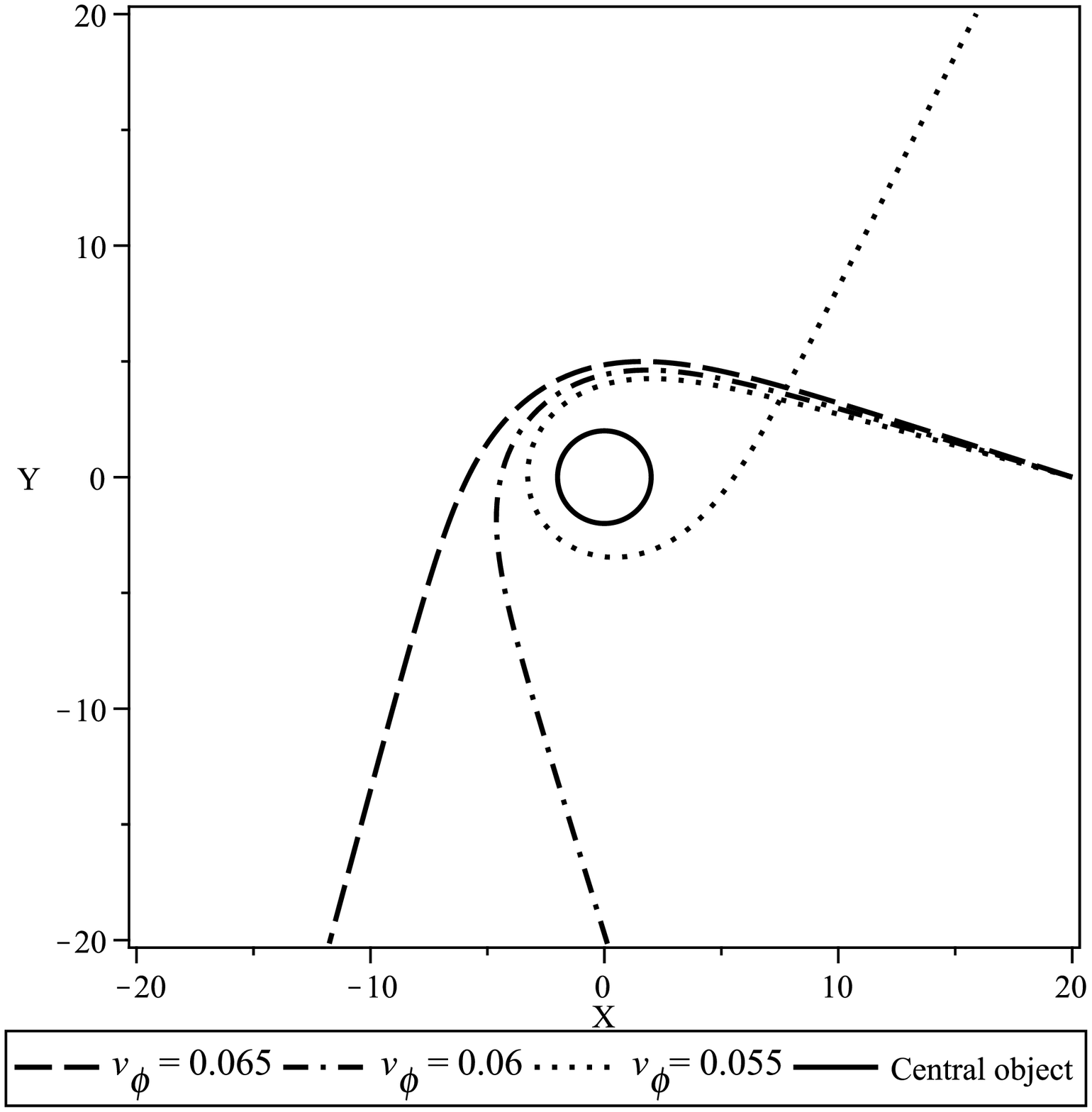} &
\includegraphics[totalheight=6cm,width=6cm]{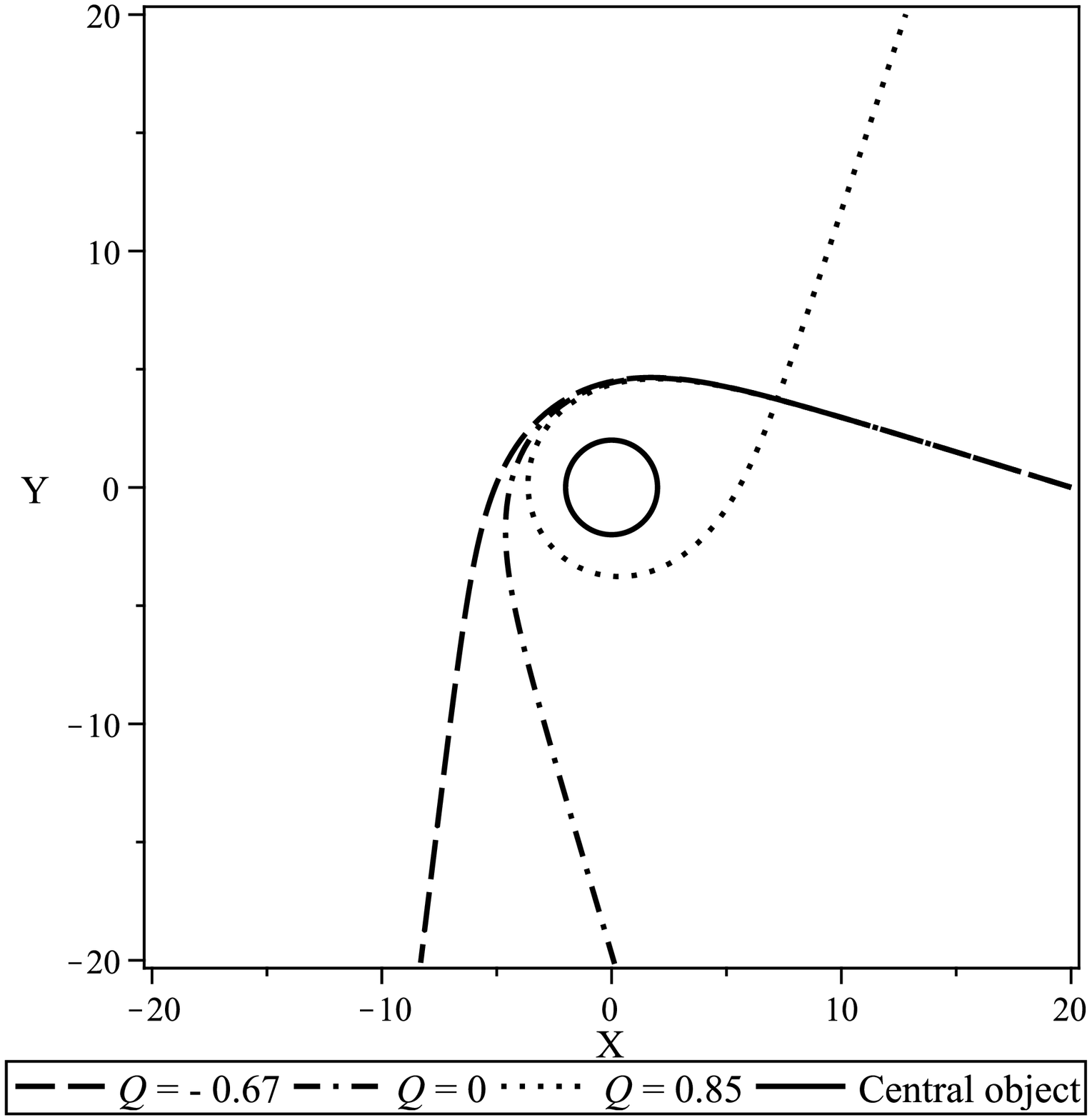}
\end{tabular}
\caption{(left) Unbound orbits. The parameters of the system are $M=1,\quad J=0,\quad Q=0,\quad r=20,\quad \phi=0,\quad \dot{r}=-4,\quad v_{\phi}=\dot{\phi}$. (right) The parameters of the system are $M=1,\quad J=0, \quad Q\neq0,\quad  r=20,\quad \phi=0,\quad \dot{r}=-4,\quad \dot{\phi}=0.06$.}\label{ploth1}
\end{figure}

\begin{figure}[t]
\centering
\begin{tabular}{lr}
\includegraphics[totalheight=6cm,width=6cm]{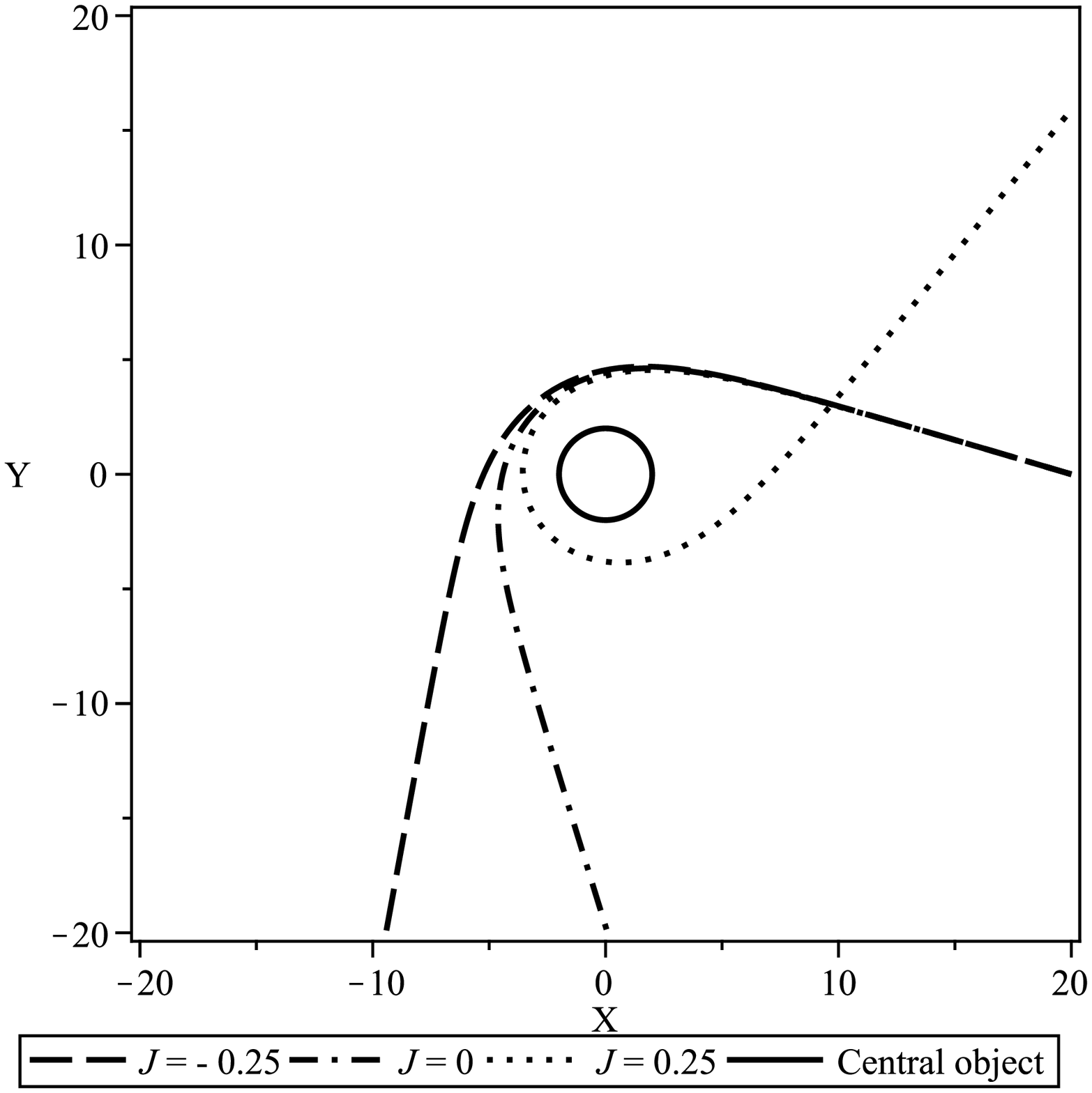} &
\includegraphics[totalheight=6cm,width=6cm]{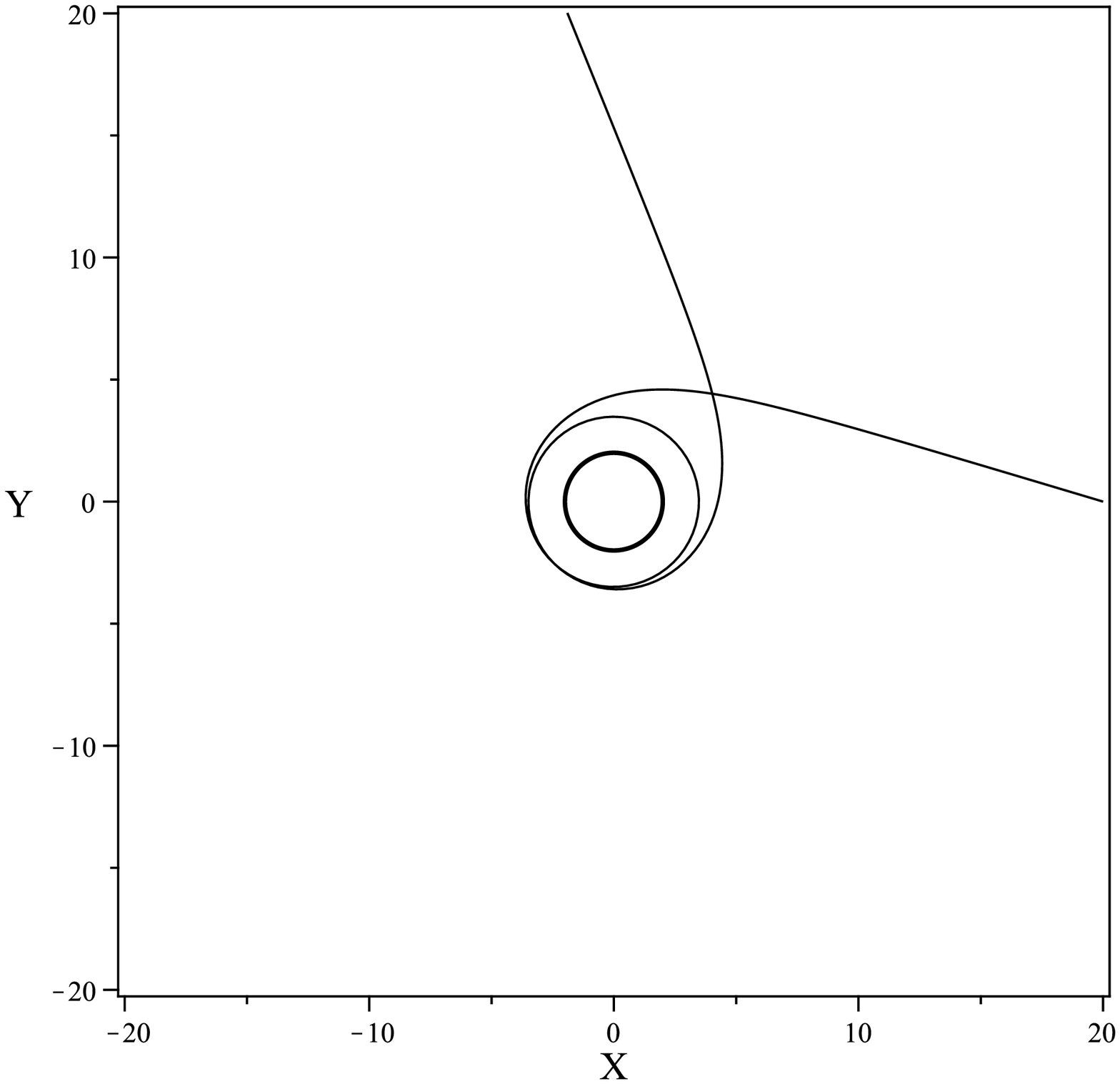}
\end{tabular}
\caption{(left) Unbound orbits. The parameters of the system are $M=1, \quad J\neq0, \quad Q=0, r=20, \quad \phi=0, \quad \dot{r}=-4, \quad \dot{\phi}=0.06$.(right) Double loop. The parameters of the system are $M=1,\quad J=0,\quad Q=0.8874, \quad r=20,\quad \phi=0,\quad \dot{r}=-4,\quad \dot{\phi}=0.06$.}\label{ploth2}
\end{figure}

It is convenient to investigate the motion of test particles numerically in the Hartle-Thorne spacetime as the full set of equations is cumbersome even for the equatorial plane. We select different values for the parameters of the source and initial conditions for test particles to consider all types of trajectories. The results of the numerical integration of timelike geodesics are shown in Figs. \ref{plotq}-\ref{ploth2}.
%%%%%%%%%%%%%%%%%%%%%%%%%%%%%%%%%%%%%%%%%%%%%%%%%%%%%%%%%%%%%%%%%%%%%%%%%%
%%%%%%%%%%%%%%%%%%%%%%%%%%%%%%%%%%%%%%%%%%%%%%%%%%%%%%%%%%%%%%%%%%%%%%%%%%
%%%%%%%%%%%%%%%%%%%%%%%%%%%%%%%%%%%%%%%%%%%%%%%%%%%%%%%%%%%%%%%%%%%%%%%%%%
%%%%%%%%%%%%%%%%%%%%%%%%%%%%%%%%%%%%%%%%%%%%%%%%%%%%%%%%%%%%%%%%%%%%%%%%%%
In Fig. \ref{plotq} (left panel), we analyze the motion of a test particle in the field of a static and deformed source. As one can see, the quadrupole parameter generates different deviations from the Schwarzschild spacetime, depending on its value and sign. The frame dragging effect is illustrated in Fig. \ref{plotq} (right panel) for a spherically symmetric source. The drag strengthens as the test particles approach the source. The influence of the frame dragging effect on the circular motion has considered in Ref.~\refcite{bini2013} for one revolution of the test particle. By analyzing the behavior of the test particles with the certain initial conditions, it is possible to select the values of $J$ and $Q$  in order to recover circular geodesics, i.e., the effects caused by the deformation of the source can be balanced by its rotation and vice-versa (see Ref.~\refcite{bini2013}).

Unbound orbits are shown in Fig \ref{ploth1} (left panel), where we change only the initial angular component of the velocity $v_{\phi}=\dot{\phi}$.  The remaining 
quantities are fixed for the sake of comparison. For different values of $Q$, we obtain the plot of  Fig. \ref{ploth1} (right panel), whereas in  Fig. \ref{ploth2} (left panel) we present
the results for different values of  $J$.

Due to this interplay between the initial conditions of the test particles and the parameters of the source, one can construct all kind of  geodesics. An example of a double loop trajectory, which is missing in classical physics, is shown in Fig. \ref{ploth2} (right panel). From here we conclude that the parameters of the geodesic motion can be used to determine the main parameters of the source such as $M$ , $J$ and $Q$.
%
%&&&&&&&&&&&&&&&&&&&&&&&&&&&&&&&&&&&&&&&&&&&&&&&&&&&&&&&&&&&&&&&&&&&&&&&&&&&&&&&&&&&&&&&&&&&&&&&&&&&&&&&&&&&&&&&&&&&&&&&&&&&&&&&&&&&&&&&&&&&&&&&&&&&&&&&&&&
\section{Conclusion}
In this work, we have explored geodesics in the Hartle-Thorne spacetime both analytically and numerically. We considered the geodesics on the equatorial plane and investigated the role of the quadrupole parameter, as well as the frame dragging effect on the motion of test particles. We investigated bounded and unbounded orbits varying the initial conditions of the test particles and the main parameters of the source. We conclude that using different combinations of both initial conditions and main parameters, one can generate many different geodesic curves.

In all our computations we used the methods of  perturbation theory. Our results have the same order of approximation as the Harte-Thorne solution. Namely, we derived the expressions for the orbital angular velocity, energy, orbital angular momentum for test particles and orbital angular velocity for photons. In turn, with the help of these expressions we obtained the radii of the innermost bound, innermost stable and photon orbits.

All the analytic expressions obtained here and in Refs.~\citen{boshkayevqpos2014} and  \citen{boshkayevqpos2015} are in agreement with the results of Refs.~\citen{A2003, 2010ApJ...714..748T, abr2004, stuchlik2014} and \citen{stuchlik2015}, if one redefines $q\rightarrow2j^2-q$.

We briefly discussed some applications of our theoretical results in the astrophysical context. In fact, the results presented in this paper can be applied to study the physics of accretion disks, the motion of test particles near the source and the epicyclic frequencies; all these aspects are of high relevance and importance in relativistic astrophysics and X-ray astronomy. For instance, using epicyclic frequencies and quasiperiodic oscillation data, one can test the strong field regime of general relativity, determine the parameters of the gravitational source and test the equations of state of compact objects. 

In a future work, we expect to apply the analytic expressions we obtained in the present work in the context of observational astrophysics. Furthermore, the investigation of the stability of the geodesics and the structure of the accretion disks are crucial to understand the physical properties of the Hartle-Thorne spacetime. We expect to perform such an analysis in a future work by applying the procedure shown, for instance, in Ref.~\refcite{quevedo2015}.

%&&&&&&&&&&&&&&&&&&&&&&&&&&&&&&&&&&&&&&&&&&&&&&&&&&&&&&&&&&&&&&&&&&&&&&&&&&&&&&&&&&&&&&&&&&&&&&&&&&&&&&&&&&&&&&&&&&&&&&&&&&&&&&&&&&&&&&&&&&&&&&&&&&&&&&&&&&

\section*{Acknowledgements}
This work was supported by the Ministry of Education and Science of the Republic of Kazakhstan Grants No. 3101/GF4 IPC-11/2015 and No. 1597/GF3 IPC-30.

\end{document}